\documentclass[11pt,a4paper]{article}
\usepackage{jheppub}
\relax

\def\bes{\begin{eqnarray}}
 \def\ees{\end{eqnarray}}
\def\be{\begin{equation}}
\def\ee{\end{equation}}
\def\bs{\begin{subequations}}
\def\es{\end{subequations}}
\newcommand{\een}{\end{subequations}}
\newcommand{\ben}{\begin{subequations}}
\newcommand{\beq}{\begin{eqalignno}}
\newcommand{\eeq}{\end{eqalignno}}

 \def\lx{\lambda}

\def\rt{{\tilde{r}}}

\def\phid{{\dot{\phi}}}

\def\ph{{\hat{p}}}
\def\xh{{\hat{x}}}

 \def\sx{\sigma}

\title{Classicalization as a tunnelling phenomenon}
\author[1]{J. Rizos,}
\author[2]{N.Tetradis}
\author[2]{and G. Tsolias}
\affiliation[1]{Department of Physics,\\ University of Ioannina, Ioannina 45110, Greece}
\affiliation[2]{Department of Physics,\\ University of Athens, Zographou 15784, Greece}

\emailAdd{irizos@uoi.gr}
\emailAdd{ntetrad@phys.uoa.gr}
\emailAdd{gtsolias@phys.uoa.gr}

\abstract{
We discuss the ``wrong"-sign DBI theory as a prototype for classicalization.
The theory lacks a UV completion and has to be defined with
a fundamental UV cutoff.
 We study a quantum-mechanical toy model with similar properties.
The model has a fundamental length scale and all physical states have momenta below the inverse of this scale.
We show that, despite the terminology, the phenomenon of classicalization is of a quantum nature.
Within the toy model it consists essentially of
tunnelling through a region that is classically forbidden. The size of this region is
proportional to the square root of the
energy and can be much larger than the fundamental length scale.
We discuss the implications for classicalization in scalar field theories.
}


\begin{document}
\maketitle

\section{Introduction}\label{intro}

The proposal of refs. \cite{dvalex1,dvpirts,dvali,dvalex2}, characterized as classicalization, concerns
the nature of high-energy scattering in certain classes of nonrenormalizable scalar field theories. It advocates that
scattering can take place at length scales much larger than the
typical scale associated with the nonrenormalizable terms in the Lagrangian. The new scale arises because the
energy of the scattering process is expected to enter in an unconventional way in the expression for the scattering
amplitude.  The scenario has been discussed in the context of various theories, such as nonlinear sigma models \cite{nonlinear}.
It has also been shown to have
very interesting collider phenomenology \cite{lhc}.
The inspiration for this scenario is taken from ultra-Planckian
scattering in gravitational theories, during which a black hole is expected to start forming at distances comparable to the
Schwarzschild radius.
This radius increases with the center-of-mass energy, and can become much larger than the fundamental scale
of the theory, set by the Planck length.
If the formation and subsequent evaporation of a black hole are viewed as parts of a scattering process, it seems reasonable
to expect
that the physical length scale determining the cross section is the Schwarzschild radius and not the Planck scale.

In the case of gravity it is difficult to support the argument with a detailed calculation. However, scalar theories
are simpler and a concrete calculation is feasible.
An idealized scattering process was proposed in refs. \cite{dvpirts,dvali} and analyzed using perturbation theory,
as a means to check the validity of the
classicalization arguments. It was subsequently studied numerically in refs. \cite{brt,rt}.
The process involves a spherical field configuration of  large radius that propagates towards
the center of symmetry.
Initially, it is assumed to have the form
\be
\phi_0(t,r)=\frac{A}{r}\exp\left[-\frac{\left(r+t-r_0\right)^2}{a^2} \right].
\label{wave} \ee
The energy density is localized within a shell of width $\sim a$ at a radius $r\simeq r_0-t$. We refer to this configuration
as a spherical wavepacket.
Using perturbation theory, it was shown in refs. \cite{dvpirts,dvali} that
the field configuration is strongly deformed when the peak of the wavepacket reaches the
classicalization radius
\be
r_*\sim L_* \left( \frac{A^2 L_*}{a}\right)^{1/3}\sim L_* \left( \sqrt{s} L_* \right)^{1/3}.
\label{clr} \ee
This can be substantially larger than the fundamental length scale $L_*$ of the theory, if the center-of-mass energy
$\sqrt{s}\sim A^2/a$ is much larger than $1/L_*$.

Behavior of this type is expected in several classes of field theories with higher-derivative terms in the Lagrangian.
As a typical example, which has been analyzed both analytically and numerically, we consider here the
Dirac-Born-Infeld (DBI) theory with Lagrangian density
\be
{\cal L}=-\frac{1}{\delta L_*^4}\sqrt{1-\delta L_*^4\left(\partial_\mu\phi \right)^2}.
\label{lagrangian} \ee
The fundamental length scale is $L_*$, as can be deduced by formally expanding the square root.
We allow for two possible signs of the first correction to the standard kinetic term by assuming that $\delta=\pm 1$.
The standard sign $\delta=1$ leads to a Lagrangian which can be viewed as providing the low-energy description
of a 3-brane in five-dimensional space-time, with the real field $\phi$ corresponding to the Nambu-Goldstone mode
arising from the breaking of translational invariance. As a result, the theory described by eq. (\ref{lagrangian})
has a well-defined UV completion. The theory with the ``wrong" sign $\delta=-1$ does not have an obvious
interpretation in the context of a more fundamental theory. In ref. \cite{dfg} it was argued that this absence
of a UV completion is linked to the fact that the theory displays behavior characteristic of classicalization.

The equation of motion of the field $\phi$ is
\be
\partial^\mu\left[ \frac{  \partial_\mu \phi}{\sqrt{1-\lx \left(\partial_\nu\phi \right)^2}}\right] =0,
\label{eom} \ee
with $\lx=\delta L^4_*$.
The momentum conjugate to the field is
\be
\pi=\frac{\phid}{\sqrt{1-\lx\phid^2+\lx (\vec{\nabla}\phi)^2}}.
\label{momconjfiel} \ee
Clearly, $\pi$ is real and finite as long as $1-\lx\phid^2+\lx (\vec{\nabla}\phi)^2>0$, a constraint that
also guarantees the reality of the Lagrangian density.
For $1+\lx (\vec{\nabla}\phi)^2\not= 0$ the above expression can be
inverted to obtain
\be
\phid=\pi \left[\frac{1+\lx (\vec{\nabla}\phi)^2}{1+\lx\pi^2}\right]^{1/2}.
\label{ppi} \ee
On the other hand, for $\lx <0$ and
$1+\lx (\vec{\nabla}\phi)^2= 0$, eq. (\ref{momconjfiel}) gives $\pi=1/\sqrt{|\lx|}$  (or $1+\lx \pi^2=0$) for all
$\dot{\phi}$. The inversion is not possible, as the whole line $1+\lx (\vec{\nabla}\phi)^2= 0$ on the
$(\dot{\phi},|\vec{\nabla}\phi|)$ plane is mapped onto the point $1+\lx\pi^2= 1+\lx (\vec{\nabla}\phi)^2= 0$
on the $(\pi,|\vec{\nabla}\phi|)$ plane.
Equivalently, it is easy to check that the Lagrangian density (\ref{lagrangian}) is singular on the
line $1+\lx (\vec{\nabla}\phi)^2= 0$, as $\partial^2{\cal L}/\partial \dot{\phi}^2$ vanishes there.

For the conjugate momentum to remain real, $1+\lx\pi^2$ and $1+\lx (\vec{\nabla}\phi)^2$
must have the same sign.
In fact, from eq. (\ref{momconjfiel}) we have
\be
\frac{1+\lx (\vec{\nabla}\phi)^2}{1+\lx\pi^2}=1-\lx\phid^2+\lx (\vec{\nabla}\phi)^2.
\label{sign} \ee
The positivity of the ratio guarantees that both the conjugate momentum (\ref{momconjfiel})
and the Lagrangian density (\ref{lagrangian}) are real.
The typical classical evolution of a field configuration  $\phi(t,\vec{x})$
in the problem of interest starts with small values of $ (\vec{\nabla}\phi)^2$
and $\pi^2$, for which  $1+\lx\pi^2>0$ and $1+\lx (\vec{\nabla}\phi)^2>0$.
For $\lx <0$ it is possible that these quantities may change sign at certain points in
space during the subsequent evolution of the collapsing wavepacket.
The conjugate momentum and the Hamiltonian density will remain real only if $1+\lx\pi^2$ and $1+\lx (\vec{\nabla}\phi)^2$ change
sign simultaneously, i.e. if the evolution goes through the point $1+\lx\pi^2= 1+\lx (\vec{\nabla}\phi)^2= 0$. As this
is a point at which the relation between $\dot{\phi}$ and $\pi$ is not invertible and a Hamiltonian density is not properly defined,
we expect the classical evolution to break down in its vicinity. We shall see in the following section that this
expectation is confirmed by the numerical integration of the equation of motion.

\begin{figure}[t]
\centering
\includegraphics[width=90mm,height=80mm]{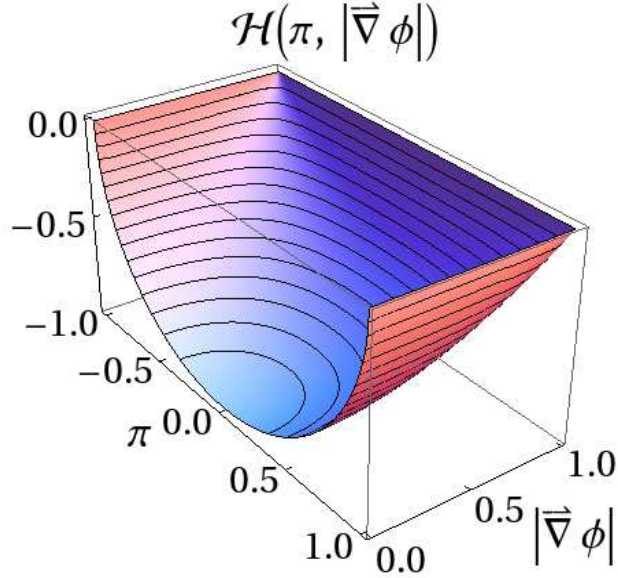}
\caption{
The Hamiltonian density for $\lx=-1$, as a function of $\pi$ and $|\vec{\nabla}\phi |$. }
\label{hamdens}
\end{figure}

Another important point concerns the range of values of the conjugate momentum allowed by the defintion (\ref{momconjfiel}).
For $\lx>0$, there is no upper bound on $\pi$. On the other hand, for $\lx<0$ and fixed $1+\lx (\vec{\nabla}\phi)^2>0$ we have
$|\pi| < 1/\sqrt{|\lx|}$, with the upper limit approached for $|\dot{\phi}|\to \infty$.

If we take into account all the above
constraints,  we are led to the conclusion that a
consistent classical evolution requires field configurations that satisfy  $1+\lx\pi^2>0$ and $1+\lx (\vec{\nabla}\phi)^2>0$
for both signs of $\lx$.
The Hamiltonian density is then given by
\be
{\cal H}=\frac{1}{\lx}\left[(1+\lx\pi^2) \left( 1+\lx (\vec{\nabla}\phi)^2\right)  \right]^{1/2}.
\label{ham} \ee
In fig. \ref{hamdens} we depict $\cal{H}$ as a function of
$\pi$ and $|\vec{\nabla}\phi|$, for $\lx=-1$. We point out that $\cal{H}$ includes a cosmological
constant term, which does not play a role in our considerations. We are interested in situations in which the
classical system starts its evolution with
$\pi$ and $|\vec{\nabla}\phi|$ sufficiently small over all space,
so that $\cal{H}$ takes values near the bottom of the ``half-bowl'' in fig. \ref{hamdens}. The subsequent
classical evolution of the system may lead, within a finite time, to configurations of the field for which $1+\lx\pi^2$
or $1+\lx (\vec{\nabla}\phi)^2 $ become zero at some points in space.
It is apparent that the classical evolution becomes pathological in
such cases. We would like to explore the possibility that a meaningful time evolution can be
defined beyond this point for the quantum system.
The discussion of the field-theoretical case poses serious technical difficulties,
as the quantized theory is perturbatively nonrenormalizable. Moreover, the implementation of the
constraints on operators such as $\pi^2$ and $ (\vec{\nabla}\phi)^2$
is not straightforward.
For these reasons, in this work
we limit our discussion within a quantum-mechanical toy model that mimics the relevant
features of the field theory.

In the following section we review the classical evolution of a spherical wavepacket within the
DBI theory. We focus on the ``wrong"-sign DBI theory, for which the classical evolution breaks down at a certain time, and discuss
the possibility that this behavior is linked to the issue of classicalization. In section \ref{quantum} we study a quantum-mechanical
system with features similar to those of the DBI theory. We show that the breakdown of the classical evolution can
be followed by a period in which the system has a purely quantum evolution, corresponding to tunnelling through a
classically forbidden region. In the concluding section we discuss the relevance of our results for the quantum
``wrong"-sign DBI theory.

\section{Classical evolution in field theory}\label{suma}

In order to make the discussion more concrete, we summarize the results of the numerical solution
of the classical equation of motion for the incoming spherical wavepacket of eq. (\ref{wave}) within the DBI theory.
When expressed in spherical coordinates, eq. (\ref{eom}) takes the form
\be
\left(1+\lx\phi_r^2 \right)\phi_{tt}-\left(1- \lx \phi_t^2 \right)\phi_{rr}
-2\lx \phi_r\phi_t \,\phi_{tr}=\frac{2\phi_r}{r}\left(1- \lx \phi_t^2+\lx\phi_r^2 \right),
\label{eomsph} \ee
where the subscripts denote partial derivatives.
In order to obtain the above equation, eq. (\ref{eom}) must be multiplied by
$\left(1- \lx \phi_t^2+\lx\phi_r^2 \right)^{3/2}$. As a result, we must assume that
$1- \lx \phi_t^2+\lx\phi_r^2 \geq 0$.
This assumption is an obvious constraint imposed by the form of the Lagrangian density (\ref{lagrangian}).
The analysis of refs. \cite{brt,rt} has
identified two new length scales in the problem.
The first scale is related to the change of type of the
quasilinear second-order partial differential equation (\ref{eomsph}).
Its discriminant is
\be
\Delta=1-\lx \phi^2_t+\lx \phi_r^2.
\label{discr} \ee
For $\Delta>0$ the equation is hyperbolic, while for
$\Delta=0$ parabolic. The case $\Delta<0$ is excluded if we require that the Lagrangian density (\ref{lagrangian})
remains real. Hyperbolic equations can support propagating waves, while parabolic ones are associated with
diffusion. We can estimate the distance $r$ at which eq. (\ref{eomsph}) may change type by evaluating $\Delta$
for the unperturbed configuration (\ref{wave}), which would be an exact solution in the absence of the nonlinear terms.
For both signs of $\lambda$, the discriminant $\Delta$ may vanish at certain points when the
wavepacket approaches a region in which $r$ is approximately given by eq. (\ref{clr}) \cite{brt}.
This seems to agree with the expectation of refs.  \cite{dvalex1,dvpirts,dvali,dvalex2} that distances
below $r_*$ cannot be probed by highly energetic incoming waves, as their propagation is impeded.

The validity of these estimates was checked in ref. \cite{rt} through
the numerical integration of eq. (\ref{eomsph}).
The solution displays different characteristics for the two signs of $\lx$. For $\lx>0$
the solution confirms the perturbative analysis of refs. \cite{dvpirts,dvali}, which predicts that
the field configuration is strongly deformed when the peak of the wavepacket reaches the
classicalization radius $r_*$. It also confirms the link with the type change of eq. (\ref{eomsph})
by demonstrating that the discriminant becomes
almost zero in certain regions of $r$.
However, the expectation that distances below $r_*$ cannot be probed is not confirmed, as the
deformed wavepacket reaches distances of the order of the fundamental length scale $L_*$.
An outgoing field configuration is also produced, but carries only a very small fraction of the
initial energy.

An important point is that the static classicalons, which were assumed to underline the
phenomenon of classicalization in ref. \cite{dvalex1}, as well as their dynamical generalizations presented
in ref. \cite{rt}, do not exist as single-valued fields over the whole space for $\lx>0$. In ref. \cite{dfg}
it was argued that this feature, as well as the existence of a UV completion of the DBI theory with $\lx >0$,
are signs that classicalization is not a feature of the DBI theory with the conventional sign.
On the other hand, the theory with the ``wrong" sign contains static and dynamical single-valued field
configurations that can be identified as classicalons \cite{dvalex1,rt,dfg}. For this reason we
focus the rest of our discussion on the case with $\lx <0$.

In ref. \cite{rt} the classical equation of motion (\ref{eomsph}) with $\lx<0$ was integrated numerically, with
the unperturbated configuration (\ref{wave}) as initial condition. It was observed that the
classical evolution breaks down before the wavepacket reaches distances of the order of the
classicalization radius. The signal for the breakdown is given by the term $1+\lx \phi_r^2$ becoming zero
at some point in space. The underlying reason has been discussed in the previous section:
The classical evolution is pathologogical  on the line $1+\lx \phi_r^2$=0.
The classical system is not shielded against such a possibility, which can occur within a finite time interval.
At first sight the underlying theory seems unphysical. However, we would like to examine whether this problem can
be resolved within the quantum theory.

Before proceeding, we point out that the radial distance at which the breakdown is expected can
be estimated by evaluating  $1+\lx \phi_r^2$ on the unperturbed configuration (\ref{wave}).
We find that this quantity vanishes at a distance
\be
\rt_*\sim L_* \left( \frac{A^2 L^2_*}{a^2}\right)^{1/2}= L_* \left( \sqrt{s} L_* \right)^{1/2}\left(\frac{L_*}{a} \right)^{1/2},
\label{newr} \ee
where we have used that $\sqrt{s}\sim A^2/a$.
If the wavepacket is assumed to have a width $a$ comparable to the fundamental scale $L_*$, we obtain
\be
\frac{\rt_*}{r_*}\sim \left( \sqrt{s} L_* \right)^{1/6},
\label{ratio} \ee
which is much larger than 1 for $\sqrt{s}\gg 1/L_*$.
The average energy density within the spherical region of radius $\rt_*$ is of order
\be
\frac{\sqrt{s}}{\rt^3_*}\sim  \frac{1}{L^{4}_*}\left( \sqrt{s} L_* \right)^{-1/2} \left(\frac{L_*}{a} \right)^{-3/2}.
\label{avdens} \ee
For $a\sim L_*$ and $\sqrt{s}\gg 1/L_*$, the density is much smaller than the typical densities $\sim L^{-4}_*$ at which
one would naively expect the classical description to fail. However, we have seen that the classical evolution can
break down already at such low densities.

\section{Evolution of the quantum-mechanical system} \label{quantum}

\subsection{One-dimensional model} \label{oned}

The purpose of this work is to develop some intuition on the physics associated with the new scale
$\rt_*$ by considering a quantum-mechanical model that displays similar behavior and can be
solved explicitly.
The model is described by
the Lagrangian
\be
L= \sqrt{1+\dot{x}^2}-\frac{1}{2}\omega^2 x^2.
\label{toy2} \ee
We continue to work in units such that $\hbar=1$.
In principle the Lagrangian is multiplied by a dimensionful energy scale $m$, which would give the
potential term its standard form in the case of a harmonic oscillator. We assume that all dimensionful
quantities are normalized with respect to this scale. This is equivalent to setting $m=1$.
Comparison of eqs. (\ref{toy2}) and (\ref{lagrangian}) demonstrates that the kinetic term has the same structure as
the DBI theory with the ``wrong" sign $\delta=-1$, when all dimensionful quantities are expressed in terms of the
fundamental length scale $L_*$. This is equivalent to setting $L_*=1$ and $\lx=-1$ in the field theory.
The quantities $1/L_*$ and $m$ play a similar role: They define the fundamental energy scales
of the respective theories.

The potential term in eq. (\ref{toy2}) has been added in order to introduce an external force that accelerates the particle and
causes its momentum to increase. In the absence of a force, the momentum remains constant and the classical
evolution is regular. The situation is different for the field theory.
In the example of an incoming spherical wavepacket,
the energy density grows because of the concentration of the energy within a diminishing central region.
The growth is achieved through the increase in magnitude of $\pi^2$ or $(\vec{\nabla}\phi)^2$.
Eventually, $1+\lx \pi^2$ or $1+\lx (\vec{\nabla}\phi)^2$ will vanish, triggering the breakdown of the classical evolution.
There is no need of an external source for the pathologies of the classical evolution to become apparent.

We also mention that the analogue of the DBI theory with the conventional sign $\delta=1$ has a kinetic term
$-\sqrt{1-\dot{x}^2}$ and describes a relativistic particle of unit mass ($m=1$).
When this is put in a quadratic potential, its classical evolution consists of
oscillations around its minimum, with a maximum velocity equal to 1. No pathologies appear.
This parallels the evolution of the incoming spherical wavepacket in the standard DBI theory, which is
regular apart from the appearance of shock fronts that have a conventional interpretation \cite{rt}.

The equation of motion resulting from the Lagrangian (\ref{toy2}) is
\be
\ddot{x}+(1+ \dot{x}^2)^{3/2}\,\omega^2 x=0.
\label{eomtoy2} \ee
The conserved energy is
\be
E=-\frac{1}{ \sqrt{1+\dot{x}^2}  }+\frac{1}{2}\omega^2 x^2 =-1 +\frac{1}{2}\omega^2 x_0^2,
\label{energytoy2} \ee
where we have assumed that initially the particle is at rest at $x=x_0$.
The conjugate momentum is $p=\dot{x}/\sqrt{1+ \dot{x}^2}$
and $|p|$ has a maximal value equal to 1, obtained for $|\dot{x}|\to \infty$. The Hamiltonian is
given by
\be
H=-\sqrt{1-p^2}+\frac{1}{2}\omega^2 x^2.
\label{hamm} \ee
Its momentum dependence is very similar to that of the Hamiltonian density (\ref{ham}) on $\pi$
for $\lx=-1$.
For $p\ll 1$ the Hamiltonian describes an ordinary harmonic oscillator. On the other hand, we must impose $|p|\leq 1$
for $H$ to remain real.

A real oscillatory solution of eq. (\ref{eomtoy2}) always exists if
$x^2_0\leq 2/\omega^2$. However, for $x^2_0>2/\omega^2$ a real solution
ceases to exist when the particle reaches the point with
$x_f^2=x^2_0-2/\omega^2$, where $\dot{x}$ diverges. This occurs within a finite time interval.
The origin of the problem can be traced
to the fact that the kinetic energy in eq. (\ref{energytoy2}) is bounded from above when considered as a function of
$\dot{x}$. The maximum occurs in the limit $|\dot{x}|\to \infty$, or $|p|\to 1$. The classical evolution cannot
satisfy the conservation of the total energy beyond this point, at which $1-p^2$ vanishes.
A similar situation occurs during the evolution of the spherical wavepacket for the DBI theory with $\lx=-1$, at the points where
$1- \pi^2$ vanishes.
The Hamiltonian density (\ref{ham}), depicted in fig. \ref{hamdens}, gives a clear
indication of the limits of the classical evolution in the field theory.
The analogy between the field theory and the quantum mechanical model persists at the quantitative level as well.
The position $x_f$ at which the classical evolution breaks down can be expressed in terms of the total energy as
$x_f=\sqrt{2E}/\omega\sim \sqrt{E}$. The energy dependence is the same as the dependence of the
scale $\rt_s$ of eq. (\ref{newr}) on the center of mass energy $\sqrt{s}$.

The question we would like to address is whether the quantum system can have a well-defined evolution even
after the classical evolution breaks down. Classicaly, the conjugate momentum is constrained by
$|p|\leq 1$ and the Hamiltonian (\ref{hamm}) remains real. We could try to eliminate this constraint for the quantum
theory. However, the form of eq. ({\ref{hamm}) indicates that states that have a nonzero projection on eigenfunctions
of the momentum operator
with eigenvalues $|p|>1$ would evolve with a complex Hamiltonian.
This type of evolution would not conserve the total energy of the system. This possibility cannot be excluded,
but would require some intuitive understanding of what happens to the dissipated energy. For this reason we shall not
consider it here. Instead, we shall assume that
the conservation of energy is guaranteed by the mechanism that underlies classicalization.
One possible implementation is to define canonical commutation relations between the
operators $\xh$ and $\ph$, but assume that the physical states of the system $|\psi\rangle$ do not
have a projection on the momentum eigenstates $|p\rangle$ with $|p|>1$. This means that in momentum space the
wavefunctions $\psi(p)=\langle p|\psi \rangle$ vanish outside the interval $|p|\leq 1$.
The continuity of the wavefunction requires that $\psi(-1)=\psi(1)=0$ in momentum space.

It is obvious that the Schr\"odinger equation can be solved most easily in momentum space.
Our choice of a quadratic potential in eq. (\ref{hamm}) is very convenient, as
the Hamiltonian takes the standard form
\be
H=-\frac{\omega^2}{2}\frac{d^2}{dp^2}+V(p),
\label{hammm} \ee
with a potential $V(p)=-\sqrt{1-p^2}$ in the interval $|p|\leq 1$. The
requirement $\psi(-1)=\psi(1)=0$ is equivalent to assuming that there are two infinite walls located at
$p=\pm 1$. In summary, the problem seems tractable: It has been reduced to solving the Schr\"odinger equation in
an infinite well with a curved bottom. The form of the potential is depicted in fig. \ref{potp}.

\begin{figure}[t]
\centering
\includegraphics[width=88mm,height=70mm]{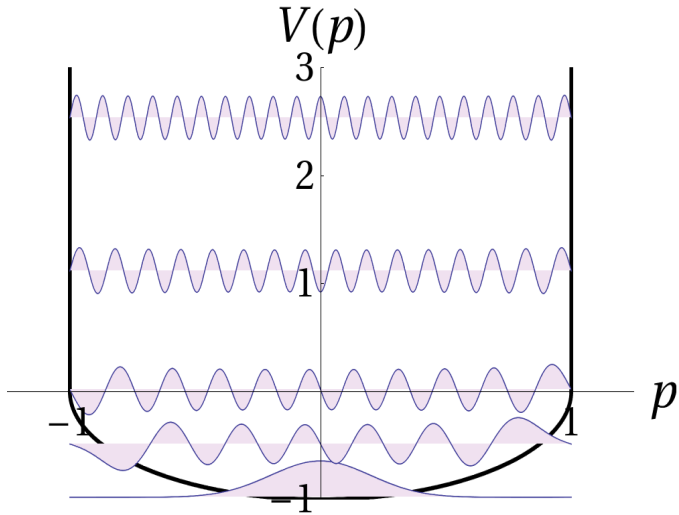}
\caption{
The potential for the solution of the Schr\"odinger equation in momentum space and the wavefunctions of several energy eigenstates.}
\label{potp}
\end{figure}

Our construction of the
quantum theory introduces a minimal length scale in the problem. The absence of momenta $|p|>1$ implies
that distances below 1 in units of the fundamental length scale $1/m$ cannot be probed. This will be apparent when we construct
wavepackets in order to reproduce the classical motion. This feature is the analogue of the
absence of a UV completion of the field-theoretical DBI model. The
similarity of the Hamiltonian density of fig. \ref{hamdens} and the bottom of the potential of fig. \ref{potp}
makes it clear that a UV cutoff is intrinsic in the DBI theory with the ``wrong" sign. As pointed out in
ref. \cite{dfg}, the theory must be considered as fundamental, without the possibility to complete it in the
UV. The classicalization argument advocates that such a theory can display nontrivial behavior at length scales
much larger than the fundamental one. Our previous discussion makes it obvious that such behavior cannot
be purely classical.

Our choice of Hamiltonian (\ref{hamm}) has introduced the new parameter $\omega$, whose
value is not fundamental for our discussion. The relevant quantity is the total
energy of the system (\ref{energytoy2}), which is determined by the combination $\omega^2 x_0^2$.
For small $\dot{x}$ the Lagrangian (\ref{toy2}) describes an ordinary harmonic oscillator
with all energy levels shifted by $-1$. We shall assume that $\omega \ll 1$, so that the separation of the energy levels is
smaller than the fundamental energy scale $m$.

In fig. \ref{potp} we depict the wavefunctions $\psi_i(p)$ of several eigenstates of the Hamiltonian (\ref{hammm})
with eigenvalues $E_i$. We have solved the Schr\"odinger equation numerically for $\omega=0.04$ for the first
60 energy eigenstates. In fig. \ref{potp} we depict the states with $i=0,11,19,30,40$, which have energies
$E_0=-0.980$, $E_{11}=-0.487$, $E_{19}=0.018$, $E_{30}=1.117$, $E_{40}=2.536$, respectively.
We depict the wavefunctions at a height equal to the corresponding energy eigenvalue. It is clear that
the low-lying states are similar to those of a harmonic oscillator of frequency $\omega=0.04$, while the
highly excited ones to those of a particle in an infinite potential well.

We would like to study the evolution of a state which, in the appropriate region, reproduces the classical
evolution of a particle described by the Lagrangian (\ref{toy2}).
This can be achieved through the construction of a wavepacket which at $t=0$ has the form
\be
\psi_0(p)=\frac{1}{\left(\sqrt{2\pi}\sigma\right)^{1/2}} \exp \left[-\frac{p^2}{4 \sigma^2} -ix_0 p \right].
\label{wavepack} \ee
It describes a particle of vanishing average momentum located at the initial position $x_0$. The dispersion of the
momenta is $\sim \sigma$, while the width of the wavepacket in position space is $\sim 1/\sigma$.
It must be noted that the wavefunction $\psi_0(p)$ does not satisfy the boundary condition
$\psi_0(-1)=\psi_0(1)=0$ that we have imposed. However, for  $\sx$ sufficiently smaller than 1, it provides an
approximation to the desired wavepacket with exponentially high accuracy.
We use $\sigma=0.3$ for the results we present in the following.
Notice also that the maximal dispersion of momenta allowed by the boundary conditions is $\sim 1$.
As a result, the width of the wavepacket in position space cannot be made smaller than $\sim 1$.
This is a manifestation of the presence of a fundamental length scale of order 1 (in units of the
scale $m$ that we discussed earlier) in the quantum-mechanical model.

\begin{figure}[t]
\includegraphics[width=\textwidth]{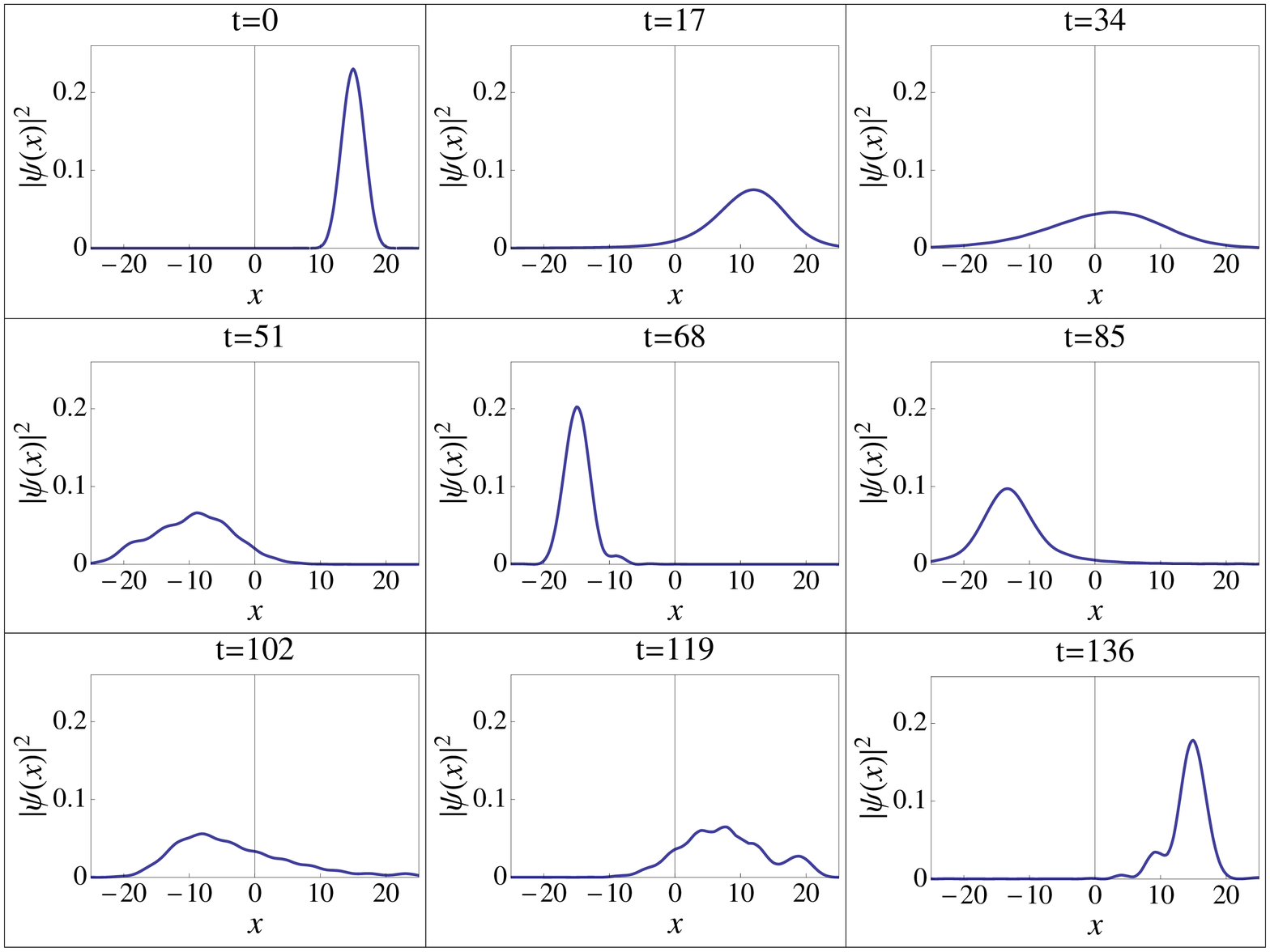}
\caption{
The evolution of a wavepacket initially located at $x_0=15$.}
\label{x015}
\end{figure}

The next step is the expansion of $\psi_0(p)$ (more precisely, of  its slight deformation that vanishes at $p=\pm 1$)
 in terms of the eigenstates of the Hamiltonian (\ref{hammm}). The boundary conditions guarantee that
this operator is self-adjoint and its eigenstates $|\psi_i\rangle$  form a complete set. We determine the projections
of  the function (\ref{wavepack}) on the first 60 such energy eigenstates. The coefficients are given by
$c_i=\langle \psi_i|\psi_0\rangle$.  In practice, we evaluate the momentum integral for each projection
within the interval $|p|\leq 1$.
The function $\sum_i c_i \psi(p)$ gives a very good approximation to $\psi_0(p)$ of eq. (\ref{wavepack})
for $|p|\leq 1$, while it vanishes for $|p|\geq 1$.
For $\sx$ sufficiently smaller than 1, the function is normalized to 1 with exponentially high accuracy.
The time evolution of the wavepacket is given by the standard expression
\be
\psi(p,t)=\sum_i c_i \psi_i(p) \exp(-iE_i t).
\label{timev} \ee
Its form in position space is obtained through a Fourier transformation. We emphasize that we have
assumed standard commutation relations for the operators $\hat{x}$, $\hat{p}$, so that the transition
from position to momentum space is standard as well. Our construction is based on the assumption that the
physical states do not have any projection on momentum eigenstates with $|p|>1$. The conservation
of energy is guaranteed by the orthonormality of the eigenstates $\psi_i(p)$, which
gives $\langle E \rangle=\sum_i |c_i|^2 E_i$.

The evolution of the wavepacket for $\omega=0.04$ and an initial position $x_0=15$ is presented in fig. \ref{x015}.
The classical motion is regular, because we have $x^2_0 < 2/\omega^2$ in this case.
We depict the propability density $|\psi(x)|^2$ as a function of position for a time interval that corresponds roughly to the period
of one oscillation. The wavepacket reproduces the classical motion to a very good approximation.
We observe some spreading of the wavepacket when it crosses the central region ($x\sim 0$), as well as some slight deformations.
However, the motion remains periodic for several oscillations with the wavepacket obtaining roughly its initial shape after each oscillation.
The period of oscillations is $T\simeq 140$, slightly longer than the time interval for which we depict the wavepacket
in fig. \ref{x015}. For sufficiently small $x_0$, the Hamiltonian (\ref{hamm}) describes a harmonic oscillator of
period $T=2\pi/\omega$. For $\omega=0.04$ the expected period of oscillations is $T\simeq 157$. The
small discrepancy with the period observed in fig. \ref{x015} is caused by the higher-order powers
of $p$ in the expansion of the square root in the Hamiltonian, which are not completely negligible for $x_0=15$.

\begin{figure}[t]
\includegraphics[width=\textwidth]{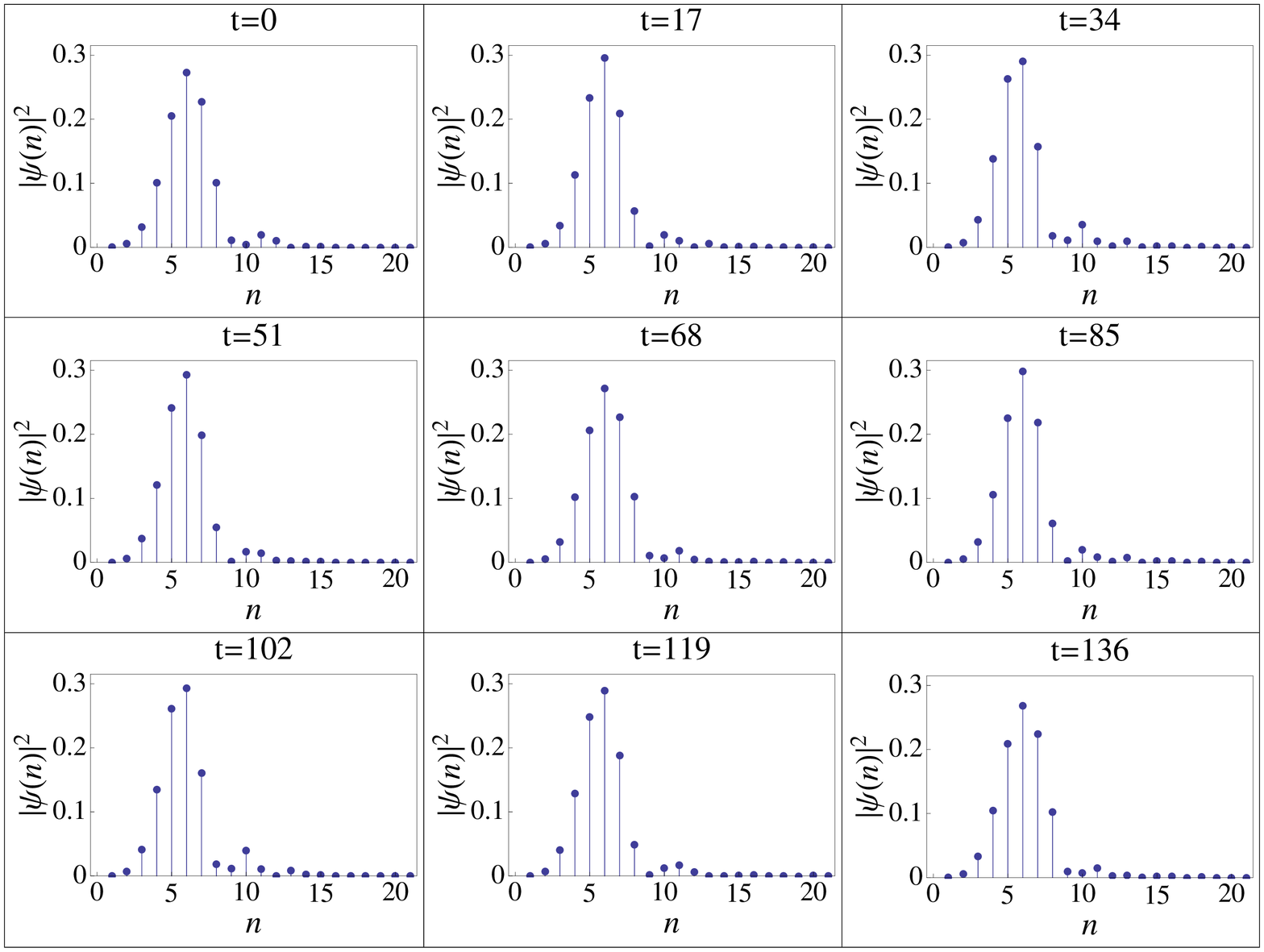}
\caption{
The decomposition of the wavepacket of fig. \ref{x015} in terms of eigenstates of the harmonic oscillator.}
\label{bathmoi15}
\end{figure}

In fig. \ref{bathmoi15} we depict the quantity $|\psi(n)|^2=|\langle n | \psi \rangle|^2$ at various times,
where the state $|n\rangle$ is
an eigenstate of a quantum harmonic oscillator of frequency $\omega$ with energy level $n$.
The purpose of this plot is to describe how the state $\psi(p,t)$ can be constructed from eigenstates of
the harmonic oscillator. In a perturbative approach to the problem one would expand the full solution
in terms of such eigenstates.
For small $x_0$ and low total energy, the wavepacket is constructed from the low-lying eigenstates of the
Hamiltonian (\ref{hammm}), which resemble those of the harmonic oscillator.
For this reason, the form of $|\psi(n)|^2$ does not change significantly with time, as can be
observed in fig. \ref{bathmoi15}.

\begin{figure}[t]
\includegraphics[width=\textwidth]{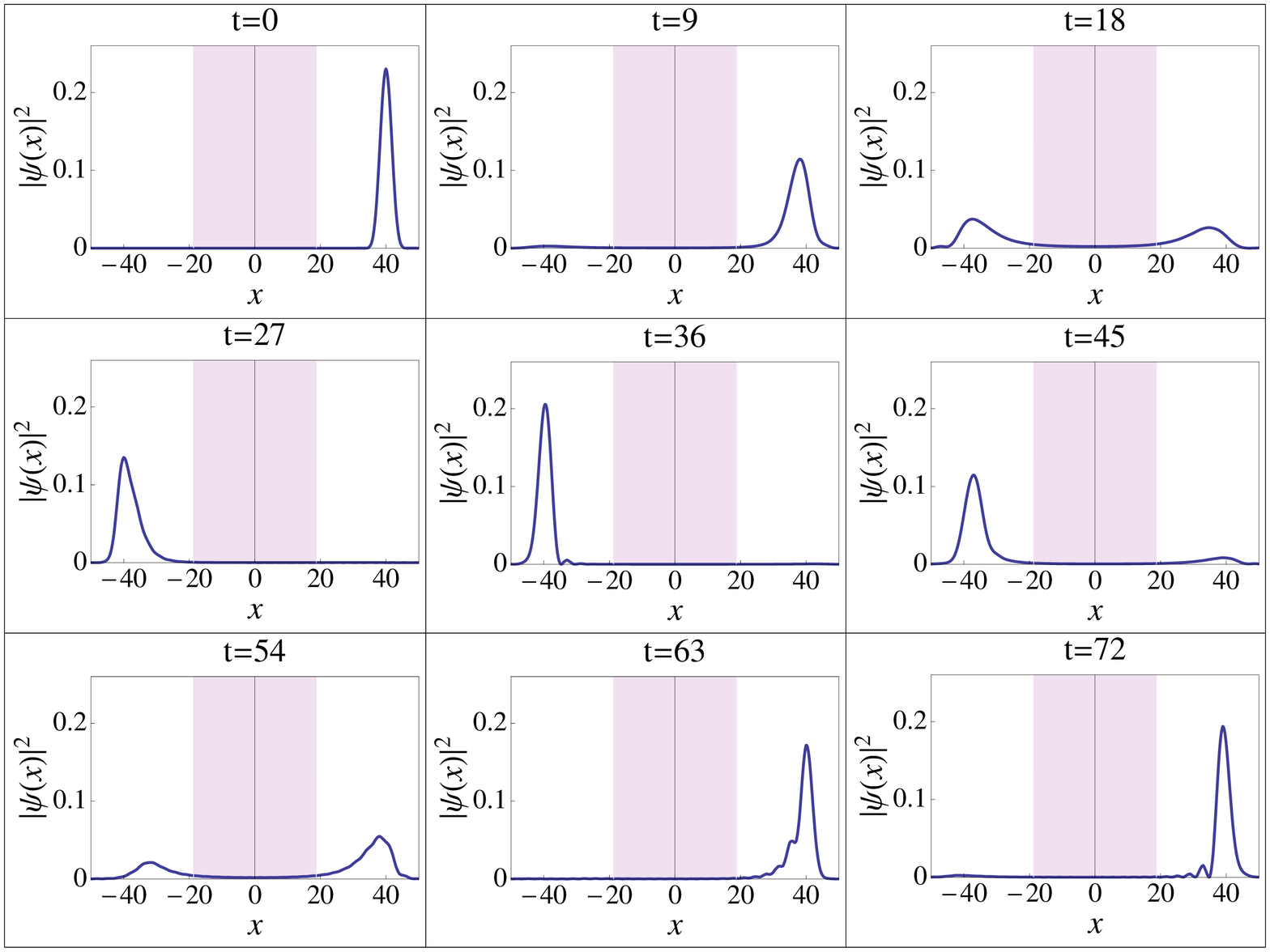}
\caption{
The evolution of a wavepacket initially located at $x_0=40$.}
\label{x040}
\end{figure}

In fig. \ref{x040} we depict the evolution of a wavepacket that starts from an initial position $x_0=40$, such that
$x^2_0> 2/\omega^2$. The classical evolution is expected to break down at $|x_f|=\sqrt{x^2_0-2/\omega^2}\simeq 18.7$.
The classicaly forbidden region correponds to the shaded area in each graph.
We observe that the wavepacket initially moves towards smaller values of $x$, retaining its
shape. However, when it approaches the right side of the classicaly forbidden region it  becomes strongly deformed.
It breaks into two pieces,
with the second one appearing on the left of the forbidden region. At later times the initial wavepacket is
reconstructed in the classically allowed region on the left.  The wavepacket displays a very intuitive physical
evolution. It tunnels through the forbidden region, similarly to the case of conventional tunnelling through
a potential barrier.  The unexpected feature in fig. \ref{x040} is the size of this region. At first sight, the presence of
a fundamental length scale $\sim 1$ in the problem, imposed by the constraint $|p|\leq 1$, would lead one to expect
that tunnelling should take place through regions of size $\sim 1$.  However, fig. \ref{x040} demonstrates that the
relevant scale is much larger and is determined by the distance at which the classical evolution breaks down.
The evolution of the quantum system guarantees the conservation of the average energy at all times, as we discussed earlier.
After its reconstruction on the left, the wavepacket evolves essentially classically until it approaches
again the forbidden region and tunnels through it.
The period of oscillations in much shorter than in the case with well-defined classical motion. We find that $T\simeq 70$.
This can be attributed to the fact that the wavepacket starts its motion at a large value of $x$ and picks up significant momentum
very quickly, so that its initial motion deviates from that of a harmonic oscillator.

\begin{figure}[t]
\includegraphics[width=\textwidth]{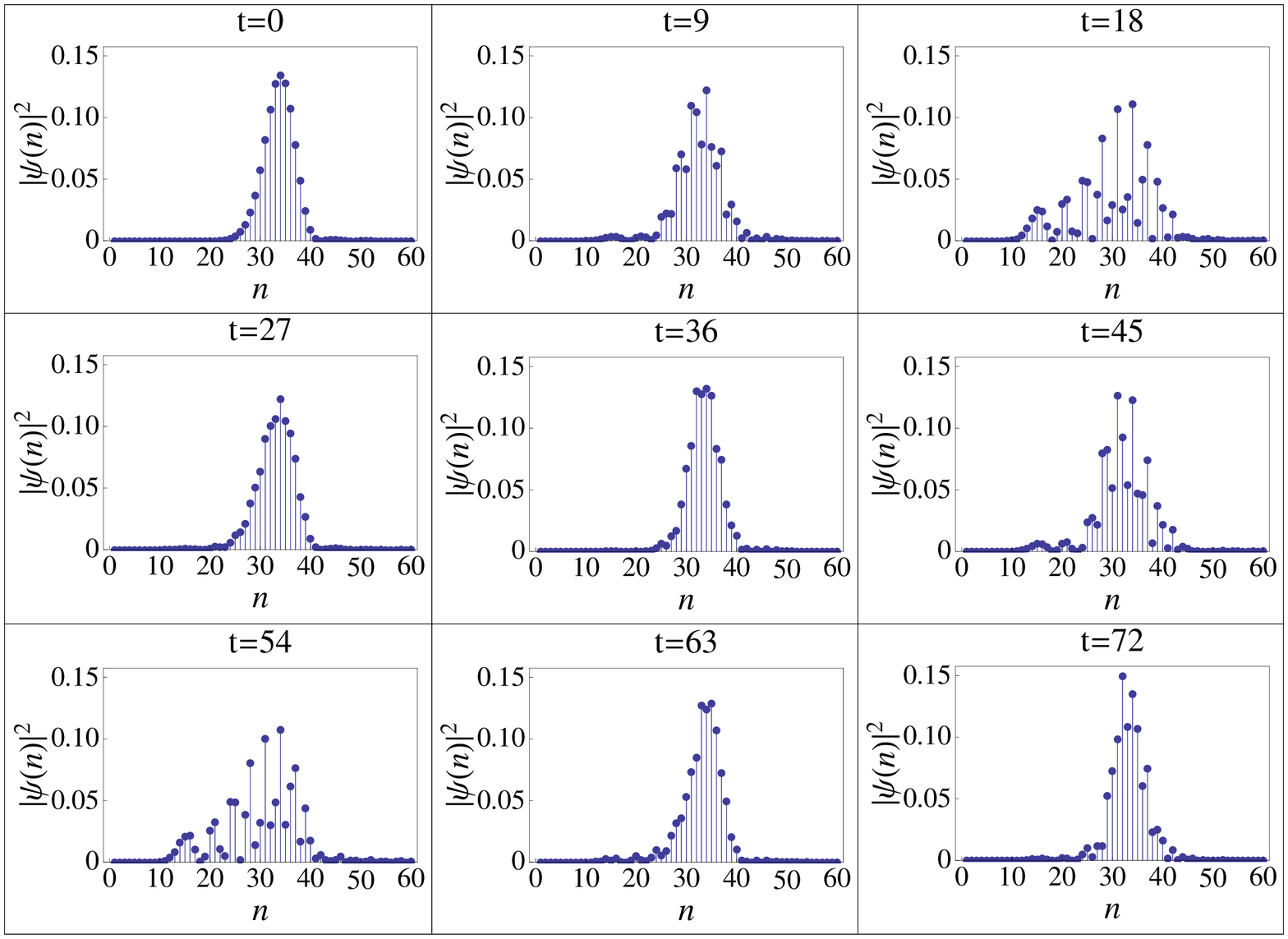}
\caption{
The decomposition of the wavepacket of fig. \ref{x040} in terms of eigenstates of the harmonic oscillator.}
\label{bathmoi40}
\end{figure}

In fig. \ref{bathmoi40} we depict the projection of the wavepacket on the eigenstates of a harmonic oscillator
of frequency $\omega=0.04$, similarly to fig. \ref{bathmoi15}. The initial wavepacket is projected on a rather narrow
range of high-energy eigenstates. Contrary to what happens in fig. \ref{bathmoi40}, the subsequent evolution
spreads the wavepacket on a much wider range of lower-energy eigenstates. The maximal spreading is obtained at the time when
the wavepacket consists mainly of two comparable parts on either side of the classically forbidden region.
The qualitative conclusion that can be drawn from fig. \ref{bathmoi40} is that, within a perturbative approach,
the transition through the classically forbidden region  could be viewed as the production of a large number of lower-energy
intermediate states.

Another point of view for the evolution of the wavepacket can be obtained by considering its form in momentum space, given by eq.
(\ref{wavepack}), and exchanging the meaning of $p$, $x$. We would then obtain a wavepacket describing a particle initially
located at the ``position" $p=0$, with initial ``momentum" $x_0$. At subsequent times, the particle will climb up the
potential $-\sqrt{1-p^2}$. If it has sufficient initial ``momentum" $x_0$, it will reflect on the wall at $p=1$ and its ``momentum"
$x_f$ will be reversed and become $-x_f$. The finite width of the wavepacket, combined with the boundary condition that forces
the wavefunction to vanish at $p=1$, causes the reflection to take place before the peak of the wavepacket reaches
the point $p=1$. The corresponding ``momentum" $x_t$ is larger than $x_f$, because the potential $-\sqrt{1-p^2}$
is an increasing function of $p^2$. This point of view explains the form of the evolution depicted in fig. \ref{x040}, in which
it is clearly visible that the transmission of propability to negative values of $x$ takes place before the peak of the wavepacket
reaches $x_f$.

It is clear from fig. \ref{x040} that, for
$x^2_0> 2/\omega^2$,
part  of the evolution of the wavepacket in position space can be interpreted as tunnelling through a classically forbidden region.
In order to justify such a
picture it is necessary to identify  quantum configurations (instantons) that could underlie this behavior.
For this purpose we switch to Euclidean time by defining $\tau=-i t$.
The model is now described by the Lagrangian
\be
L_E=- \sqrt{1-\left( \frac{dx}{d\tau} \right)^2}+\frac{1}{2}\omega^2 x^2,
\label{toy2e} \ee
which has a very physical interpretation: It describes a relativistic particle moving in an
inverted quadratic potential. We need the solution of the effective equation of motion for initial conditions that match the
final stage of the classical evolution. As during that stage the particle moves towards the minimum of the potential, it is
natural that in the Euclidean picture the particle initially moves towards the top of the inverted potential.
However, the Euclidean velocity cannot exceed a maximal value equal to 1. The consistency of the two pictures
leads us to the conclusion that tunnelling becomes
relevant even before the classical velocity diverges at $x_f^2=x^2_0-2/\omega^2$. As soon as the
velocity becomes
sufficiently close to 1, the Euclidean equation of motion has a solution that describes a relativistic particle with sufficient
energy to climb up one side of the inverted quadratic potential and roll down the other.
An instanton would be a solution of the equation of motion resulting from the Lagrangian (\ref{toy2e}), which would
start from some initial position $x_t$ and end at $-x_t$.
A rough estimate for $x_t$ can be obtained by
setting $\dot{x}\simeq 1$ in eq. (\ref{energytoy2}).
We obtain $x^2_t=x^2_0-1/\omega^2$. For $x_0=40$ and $\omega=0.04$ we find
$x_t\simeq 31.2$. This explains why the tunnelling in fig. \ref{x040} starts before the wavepacket reaches the
value $x_f\simeq 18.7$. It is also apparent that the tunnelling takes place between points that are in very good
agreement with our estimate for $x_t$.

We would like to emphasize that the above interpretation is intuitive rather
than rigorous. The particularity of the physical system we are studying requires the simultaneous use of the
Minkowskian and Euclidean pictures.
The configuration that we characterize as instanton does not correspond to a transition between two minima of the
potential, as in the conventional picture. Moreover, it remains to be shown how to match formally the formulations with
real and imaginary time. Despite these open questions, the picture of tunnelling through the classically forbidden region
is clearly supported by  the exact solution of the quantum evolution, as depicted in fig. \ref{x040}.

\begin{figure}[t]
\includegraphics[width=\textwidth]{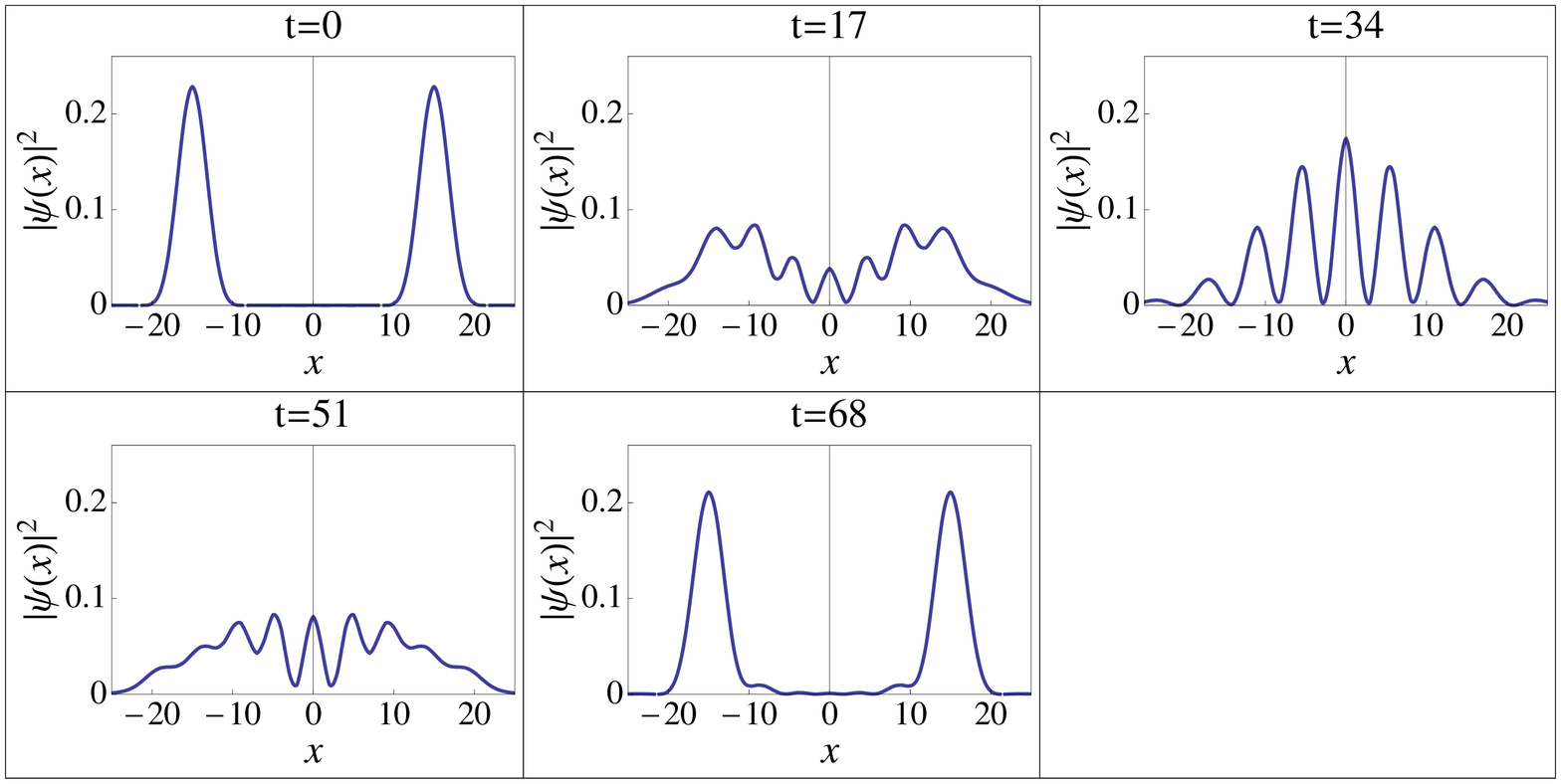}
\caption{
The evolution of two wavepackets initially located at $x_0=\pm15$.}
\label{2x015}
\end{figure}

\subsection{Interaction of wavepackets}

It is interesting to consider the case of two wavepackets initially located at symmetric points around the origin in position
space.
The relevant wavefunction can be constructed as the Fourier transform of the superposition of two configurations of the type given
by eq. (\ref{wavepack}) with opposite values of $x_0$. The wavepackets are expected to move towards the
minimum of the quadratic potential at the origin and interact during the time that they overlap.
Their evolution can be obtained by computing the projection of the initial wavefunction on the eigenstates of the
Hamiltonian (\ref{hammm}), exactly as we did for the single wavepacket. The wavefunction at all subsequent times is given by
eq. (\ref{timev}).

In fig. \ref{2x015} we depict the evolution of two wavepackets initially located at $x_0=\pm 15$ with vanishing momentum.
We have normalized the total wavefunction to 2.
Under the influence of the potential the wavepackets move towards the origin where they overlap, producing
very strong interference effects. Eventually each one moves to the side opposite to where it started and stops at
a point symmetric to its initial position. The process is then repeated periodically.

The situation looks completely different if the wavepackets start at sufficiently large values of $|x|$.
In fig. \ref{2x040} they start at $x_0=\pm 40$, so that there is a classically forbidden region corresponding to
$x^2<x_f^2=x_0^2-2/\omega^2$. The initial evolution of the wavepackets brings them closer to the minimum of the
potential. They subsequently tunnel through the forbidden region and reappear on the opposite side. There is a time interval around
$t=18$ when each wavepacket would be split between the two sides of the classically forbidden region (see fig. \ref{x040}). Their overlap
results in the strong interference effects observed in fig. \ref{2x040}.
At no stage of the evolution there is significant probability density within the classically forbidden region.

It is tempting to interpret the process depicted in fig. \ref{2x040} as scattering, characterized
by a length scale of the order of $x_f=18.7$. However, there are problems with such an interpretation. The tunnelling
through the central region is a result of the total energy of each incoming wavepacket
being larger than the maximal
kinetic energy it can accomodate, and not of its interaction with the other wavepacket.
The kinetic energy increases as each wavepacket moves down the external potential until
the classical evolution cannot be sustained any more. There are interference effects when the
wavepackets overlap, but this is a secondary phenomenon.

\begin{figure}[t]
\includegraphics[width=\textwidth]{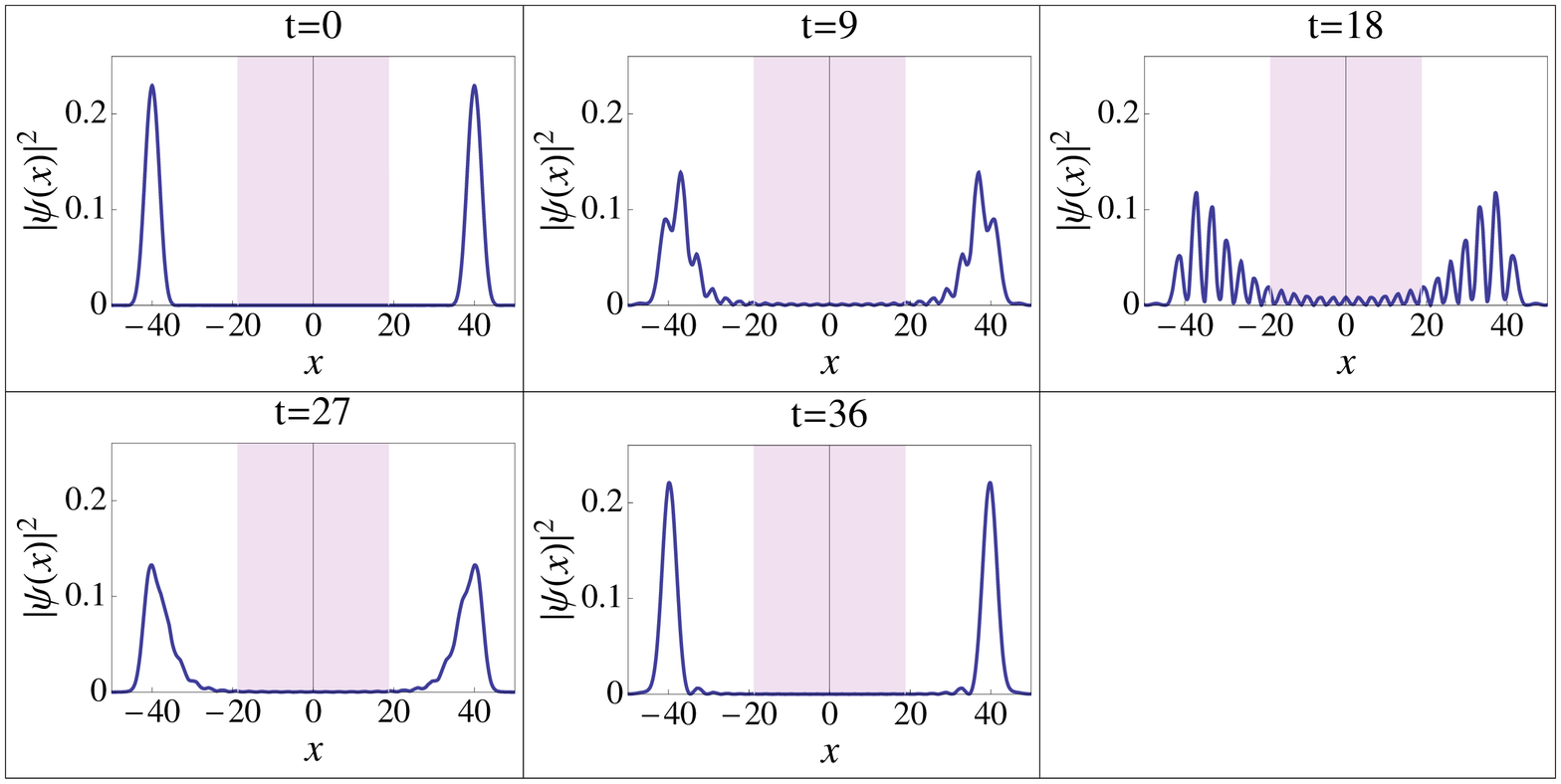}
\caption{
The evolution of two wavepackets initially located at $x_0=\pm40$.}
\label{2x040}
\end{figure}

On the other hand, the similarity with the classical evolution in the ``wrong''-sign DBI field theory, summarized in
section \ref{suma}, provides some important hints. For the field theory the total energy is conserved, but there is no external potential.
The concentration of the total energy within a continuously decreasing region forces the energy density to
increase. The simultanous increase of the values of $\pi^2$ and $(\vec{\nabla}\phi)^2$ results in the breakdown of the
classical evolution. Inspection of the Hamiltonian denisty
(\ref{ham}) for $\lx<0$ indicates that the UV cutoff must implemented in such a way that it constrains the expectation values of
operators such as
$\pi^2$ and $(\vec{\nabla}\phi)^2$ to remain smaller than $1/|\lambda|=L^{-4}_*$. The quantum-mechanical scenario that
we discussed implies that the
quantum field will avoid the classically excluded region in a similar fashion, by tunnelling through it.
In a scattering process within the quantum field theory, the energy that is concentrated within a small region around the
center of the collision is not provided by an external source, as in the quantum-mechanical model, but corresponds to the
kinetic energy of the initial particles. As a result, the quantum process, with the inclusion of tunnelling, would be a true
scattering event.

\subsection{Three-dimensional model} \label{threed}

In order to develop intution on the problem of the collapsing spherical shell in field theory, we also
consider the three-dimensional quantum-mechanical system with a Lagrangian of the form
\be
L= \sqrt{1+\dot{\vec{x}}^2}-\frac{1}{2}\omega^2 \vec{x}^2.
\label{toy3} \ee
If the wavefunction is assumed to be spherically symmetric, it provides the framework closest to the
field theoretical problem we discussed in section \ref{suma}.
The steps in the analysis of this system are completely analogous to the ones we followed in the
previous section for the one-dimensional system.
The conjugate momentum is $\vec{p}=\dot{\vec{x}}/\sqrt{1+ \dot{\vec{x}}^2}$
and $|\vec{p}|$ has a maximal value equal to 1, obtained for $|\dot{\vec{x}}|\to \infty$. The Hamiltonian is
given by
\be
H=-\sqrt{1-\vec{p}^2}+\frac{1}{2}\omega^2 \vec{x}^2.
\label{hamm3d} \ee
Under the assumption of spherical symmetry, the Hamiltonian operator in momentum space takes the form
\be
H=-\frac{\omega^2}{2}\frac{1}{p}\frac{d^2}{dp^2}p+V(p),
\label{ham3d} \ee
with a potential $V(p)=-\sqrt{1-p^2}$ in the interval $0\leq p\leq 1$. The wavefunction $\psi(p)$ satisfies
$\psi(1)=0$ and $d\psi(0)/dp=0$.
The energy eigenstates $\psi_i(p)$
are determined numerically, similarly to the previous subsection. The low-energy ones resemble the
eigenstates of the spherical harmonic oscillator, while the high-energy ones the eigenstates of an infinite spherical well.

\begin{figure}[t]
\includegraphics[width=\textwidth]{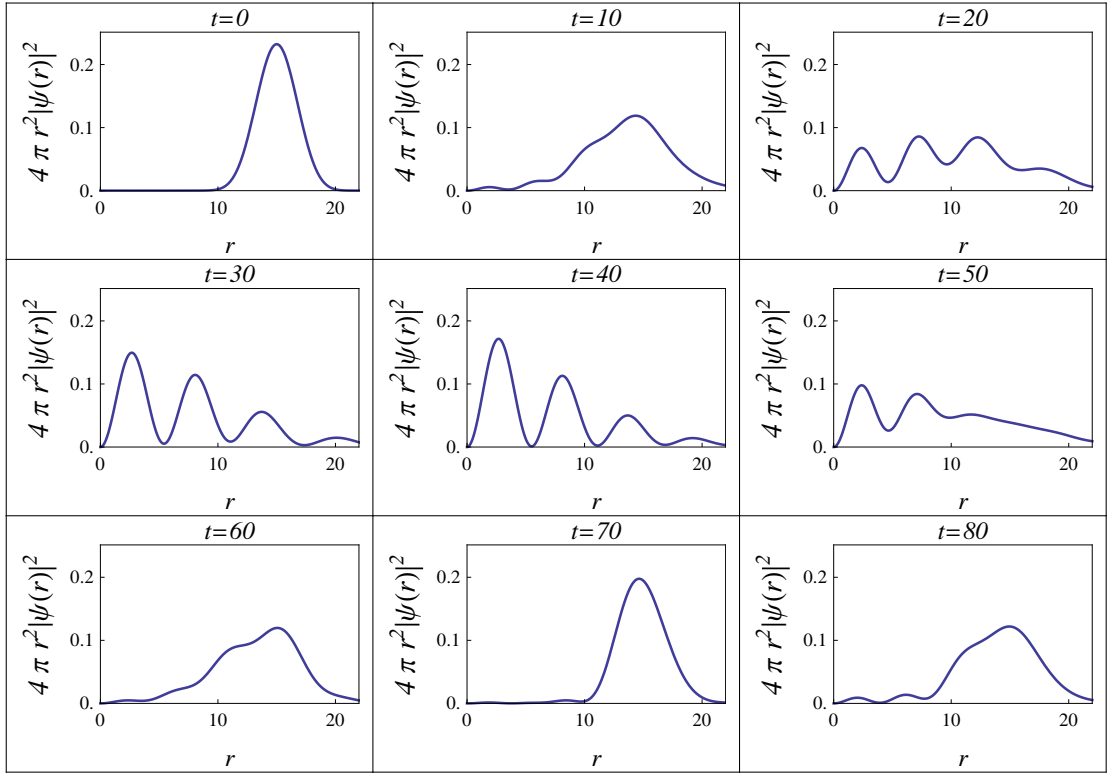}
\caption{
The evolution of a spherical wavepacket initially located at $r_0=15$.}
\label{r015-3d}
\end{figure}

We study the evolution of a spherical wavepacket with the initial form
\be
\psi_0(p)=\frac{1}{\left(\sqrt{2\pi^3}\sigma\right)^{1/2}} \frac{1}{p}\exp \left[-\frac{p^2}{4 \sigma^2}  \right]\sin(p r_0)
\label{wavepack3d} \ee
in momentum space. For $\sigma r_0 \gg 1$, its form in position space is
\be
\psi_0(r)=\frac{\sigma^{1/2}}{(2 \pi)^{3/4}} \frac{1}{r}\exp \left[-(r-r_0)^2\sigma^2  \right].
\label{wavepack3dr} \ee
We assume that $\sigma$ is sufficiently smaller than 1 for the function (\ref{wavepack3d}) to be exponentially small for
$p >1$.
We project $\psi_0(p)$ onto the eigenstates $\psi_i(p)$ and obtain the time evolution of the wavepacket according to
eq. (\ref{timev}).

\begin{figure}[t]
\includegraphics[width=\textwidth]{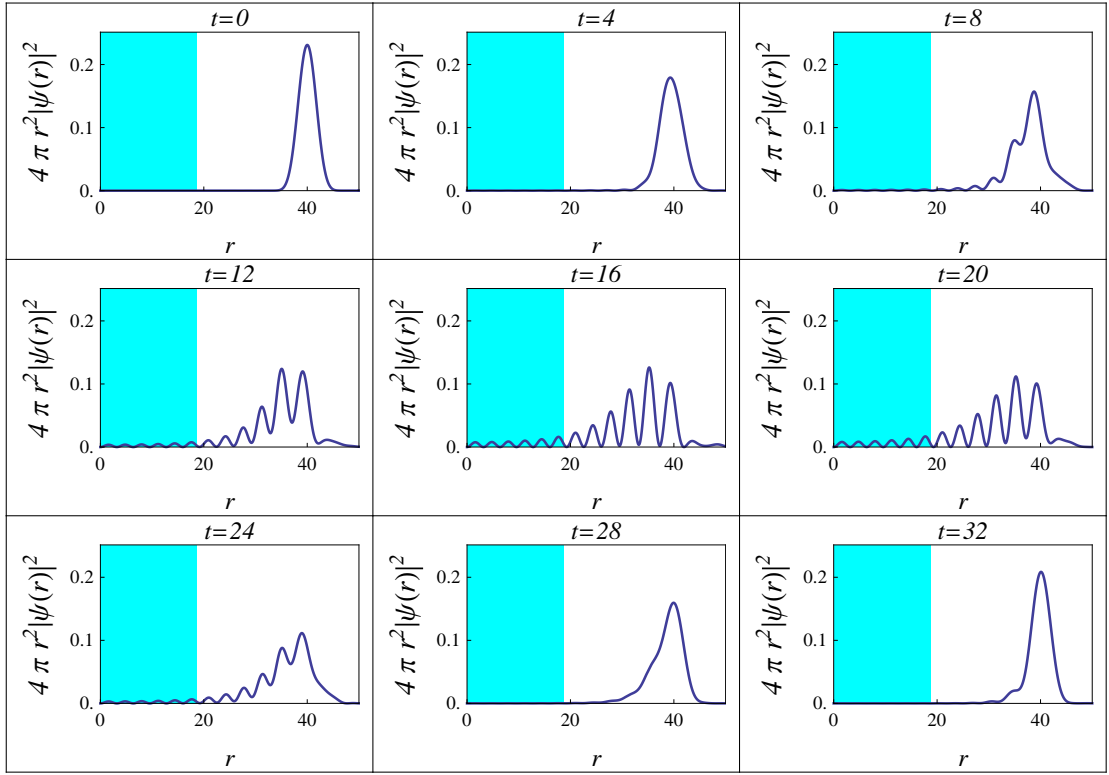}
\caption{
The evolution of a spherical wavepacket initially located at $r_0=40$.}
\label{r040-3d}
\end{figure}

In fig. \ref{r015-3d} we present the evolution of a wavepacket with $\sigma=0.3$ and $r_0=15$. We use
$\omega=0.04$, as in the previous subsection. We depict the quantity
$4 \pi r^2|\psi(r)|^2$, which provides the propability density at radius $r$.
We observe the motion of the wavepacket towards the central region and its subsequent evolution towards
its initial form. Each part of the spherical wavepacket passes through the origin and
evolves towards a point symmetric to its initial position.
Strong interference effects are visible when various parts of the spherical wavepacket converge
at $r=0$.

The evolution for $r_0=40$ is presented in fig. \ref{r040-3d}. The shaded region depicts the classically
forbidden region, corresponding to $r<r_f$ with $r_f^2=r_0^2-2/\omega^2$. It is apparent that the bulk of
the wavepacket cannot enter the region with $r<r_f$. Strong interference effects are
visible again, in the region where the incoming and outgoing parts of the wavepacket overlap.
The evolution of the wavepacket
can be viewed either as reflection or as tunnelling through the central region. The intuition developed in the previous
subsection through the study of the one-dimensional system favors the second interpretation.
Another indication is the presence of an instanton configuration in the Euclidean version of the model, in complete analogy
to the one-dimensional model that we discussed in the previous subsection. It is a spherically symmetric
configuration whose profile corresponds to the motion of a relativistic particle, which starts with an
initial velocity close to 1, climbs up the
inverted potential and eventually stops at its top.
Exactly as in the previous subsection, we can estimate that tunnelling becomes relevant
when the wavepacket reaches a radius $r_t$ given by $r^2_t=r^2_0-1/\omega^2$.
For the parameter values that we have used, we obtain $r_t\simeq 31.2$.
The evolution depicted in fig. \ref{r040-3d} is in very good quantitative agreement with this estimate.

\section{Conclusions} \label{conclu}

The evolution of the quantum-mechanical system that we analyzed displays the characteristics attributed to classicalization.
It is a unique example in which a complete calculation has produced the conjectured  behavior. For this reason we expect that
it captures features that are relevant for the field-theoretical models as well.
Our analysis provides an explicit demonstration of how a new scattering scale may appear in theories
with nontrivial kinetic terms.
It also reveals two properties of the underlying mechanism:
\begin{itemize}
\item
The theory has a fundamental length scale and all physical states have momenta below the inverse of this scale.
In the field-theoretical context, we expect that theories with similar behavior would not have a UV completion.
This point has been emphasized in ref. \cite{dfg} as well.
\item
Despite the terminology, the phenomenon of classicalization is of a quantum nature.
Within the toy model we studied, it consists essentially of
tunnelling through a region that is classically forbidden.
The size of this region is proportional to the square root of the
energy and can be much larger than the fundamental length scale.
\end{itemize}

It seems very likely that classicalization is not a feature of all higher-derivative theories.
Such theories are expected to display nonlocal behavior when the scattering takes place at energies larger
than inverse of their fundamental length scale. For example, in the context of the
conventional DBI field theory the scattering starts already at what has been termed
the classicalization radius \cite{dvpirts,rt}. However, in this
case length
scales smaller than the classicalization radius can be probed in a scattering experiment \cite{rt}. The corresponding
quantum-mechanical model has similar properties.
On the other hand,
the ``wrong''-sign DBI theory, with its very particular structure, has the potential to display classicalization.
We demonstrated that, within the corresponding quantum-mechanical model, high-energy scattering cannot probe
distances below a new energy-related length scale, much larger than the fundamental one.

Our analysis of the quantum-mechanical analogue of the ``wrong"-sign DBI theory implies that the
breakdown of the classical evolution in the field theory may be associated with the transition to a quantum regime in which
tunnelling plays a significant role.
In this respect, it is worth pointing out the presence of instanton-like configurations within the Euclidean
field theory. If we switch to Euclidean time $\tau=-i t$, the action of the DBI theory with $\lx=-1$
has a saddle point which is a solution of the equation
\be
-\left(1-\phi_r^2 \right)\phi_{\tau \tau}-\left(1- \phi_\tau^2 \right)\phi_{rr}
-2 \phi_r\phi_\tau \,\phi_{\tau r}=\frac{2\phi_r}{r}\left(1-  \phi_\tau^2-\phi_r^2 \right).
\label{eomsphe} \ee
This is nothing but eq. (\ref{eomsph}), when dimensionful quantities are normalized with respect to
$L_*$, so that $\lx=-1$, and the Euclidean time is used.
It has a solution of the form $\phi(z)$, with $z=\sqrt{\tau^2+r^2}$, and $\phi'(z)=h(z)$,  with
\be
h(z)=\pm \frac{1}{\sqrt{c^2 z^6+1}}.
\label{instanton} \ee
This configuration has a singularity in the second derivatives around the origin at $\tau=r=0$. It is actually
a solution of eq. (\ref{eom}) with a r.h.s.  that includes a source term $\sim \delta(z)$.
However, it must be kept in mind that the theory has an intrinsic cutoff scale $\sim 1$ in units of $L_*$, so that
singularities appearing at smaller length scales are unphysical. Eq. (\ref{instanton})
describes a field configuration with 
an effective range $\sim c^{-1/3}$.
When viewed as a function of $\tau$, the field
vanishes at $\tau\to -\infty$, becomes nonzero within a spherically symmetric region of space of radius $\sim c^{-1/3}$
at $\tau \sim 0$, and vanishes again at $\tau \to \infty$.
The configuration (\ref{instanton}) can also be identified with the static classicalon of
ref. \cite{dvalex1} in the case of four spatial dimensions (with $x_4=\tau$).

It remains to be seen through an explicit calculation
how the phenomena we described in this work can appear within a consistent
quantum field theory.  The nonlocal behavior we observed is a consequence of the presence of
unconventional kinetic terms in the Lagrangian and the lack of a UV completion. These properties of the theory may also generate
undesirable features. For example,
the ``wrong"-sign DBI theory displays superluminality on certain nontrivial backgrounds \cite{inflation1,adams}, even though it
is not clear how serious a pathology this is \cite{babichev,dfg}.
On the other hand, the theory also predicts interesting novel phenomena, such as a large tensor-to-scalar ratio in the
context of inflation \cite{inflation1}, or the possibility to emit signals from inside a black hole \cite{bh1,bh2}.
Understanding the essence of classicalization in scalar field theories
requires further study in the direction we explored in this work.

\section*{Acknowledgments}
We would like to thank N. Brouzakis, F. Diakonos, G. Dvali, C. Germani, C. Gomez, F. Hadjioannou and K. Tamvakis 
for useful discussions.
This research has been supported in part by
the ITN network ``UNILHC'' (PITN-GA-2009-237920).
This research has been co-financed by the European Union (European Social Fund – ESF) and Greek national
funds through the Operational Program ``Education and Lifelong Learning" of the National Strategic Reference
Framework (NSRF) - Research Funding Program: ``THALIS. Investing in the society of knowledge through the
European Social Fund".

\end{document}